# Carboneyane: a Nodal Line Topological Carbon with $sp-sp^2-sp^3$ Chemical Bonds

Jing-Yang You[1], Xing-Yu Ma[1], Zhen Zhang[1], Kuan-Rong Hao[1], Qing-Bo Yan[2], Xian-Lei Sheng[3*], Gang Su[1,4*]

[1] School of Physical Sciences, University of Chinese Academy of Sciences, Beijing 100049, China

[2] College of Materials Science and Opto-Electronic Technology, University of Chinese Academy of Sciences, Beijing 100049, China

[3] Department of Physics, Key Laboratory of Micro-nano Measurement-Manipulation and Physics (Ministry of Education), Beihang University, Beijing 100191, China

[4] Kavli Institute for Theoretical Sciences, and CAS Center for Excellence in Topological Quantum Computation, University of Chinese Academy of Sciences, Beijing 100190, China

**Abstract**

A structurally stable carbon allotrope with plentiful topological properties is predicted by means of first-principles calculations. This novel carbon allotrope possesses the simple space group C2/m, and contains simultaneously $sp$, $sp^2$ and $sp^3$ hybridized bonds in one structure, which is thus coined as carboneyane. The calculations on geometrical, vibrational, and electronic properties reveal that carboneyane, with good ductility and a much lower density 1.43 g/cm$^3$, is a topological metal with a pair of nodal lines traversing the whole Brillouin zone, such that they can only be annihilated in a pair when symmetry is preserved. The symmetry and topological protections of the nodal lines as well as the associated surface states are discussed. By comparing its x-ray diffraction pattern with experimental results, we find that three peaks of carboneyane meet well with the detonation soot. On account of the fluffy structure, carboneyane is shown to have potential applications in areas of storage, adsorption and electrode materials.

*Corresponding author. E-mail: xlsheng@buaa.edu.cn (Xian-Lei Sheng); gsu@ucas.ac.cn (Gang Su)

## 1. Introduction

Carbon is an extremely versatile element in the periodic table, as it can form *sp*, *sp*$^2$ and *sp*$^3$ hybridized chemical bonds, which thus has a strong ability to bind itself with other elements to generate countless organic compounds with chemical and biological diversity, resulting in the present colorful world. Since 1980s, several new allotropes of elemental carbon, including fullerenes [1], carbon nanotubes [2], and graphene [3], etc., have been synthesized, which inspired numerous attempts to find new structures of carbon in the past decades. According to the Samara Carbon Allotrope Database (SACADA) [4], except the above-mentioned celebrated structures, more than 500 carbon allotropes have been proposed, among which only a few carbon allotropes such as one-dimensional *sp*-carbyne [5], two-dimensional *sp*-*sp*$^2$-graphdiyne [6], and three-dimensional *sp*$^3$ T-carbon [7,8], etc. were synthesized in experiments. In addition, there are many carbon allotropes with e.g. *sp*$^2$ [9], *sp*$^3$ [10], *sp*-*sp*$^3$ [11,12], *sp*$^2$-*sp*$^3$ [13,14], hybridized bonds have been predicted. However, there is no reported carbon allotrope that contains simultaneously all *sp*, *sp*$^2$ and *sp*$^3$ hybridized chemical bonds in one structure that may display very intriguing physical and chemical properties.

In this paper, we propose a new carbon allotrope that contains three types of hybridized chemical bonds *sp*, *sp*$^2$ and *sp*$^3$ of carbon atoms in one structure, thus coined as carboneyane, which is energetically, kinetically and dynamically stable and with a much lower density 1.43 g/cm$^3$. It has a simple space group C2/m, and is a topological metal with a pair of nodal lines traversing the whole Brillouin zone (BZ). In addition, because of its fluffy structure, we find that carboneyane could be a promising storage material for Li, K and Mg atoms, and a good anode material for lithium-ion and magnesium-ion batteries, revealing that it might have wide applications.

## 2. Computation methods

The first-principles calculations were done with the Vienna ab initio simulation package (VASP) using the projector augmented wave (PAW) method in the framework of density functional theory (DFT) [15,16,17]. The electron exchange-correlation functional was described by the generalized gradient approximation (GGA) in the form proposed by Perdew, Burke, and Ernzerhof (PBE) [18]. The structure relaxation considering both atomic positions and lattice vectors was performed by the conjugate gradient (CG) scheme until the maximum force on each atom was less than 0.001 eV/Å, and the total energy was converged to 10$^{-6}$ eV with Gaussian smearing method. The energy cutoff of the plane waves was chosen as 520 eV. The BZ integration was sampled by using a 13×13×13 G-centered Monkhorst-Pack grid for the calculations of relaxation and electronic structures. The phonon frequencies were calculated using a finite displacement approach as implemented in the PHONOPY code [19], in which a 2×2×2 supercell containing 64 carbon atoms and a displacement of 0.01 Å from the equilibrium atomic positions are employed. An effective tight-binding Hamiltonian constructed from the maximally localized Wannier functions (MLWF) was used to investigate the surface states [20,21]. The iterative Green function method [22] was used with the package WannierTools [23].

## 3. Results

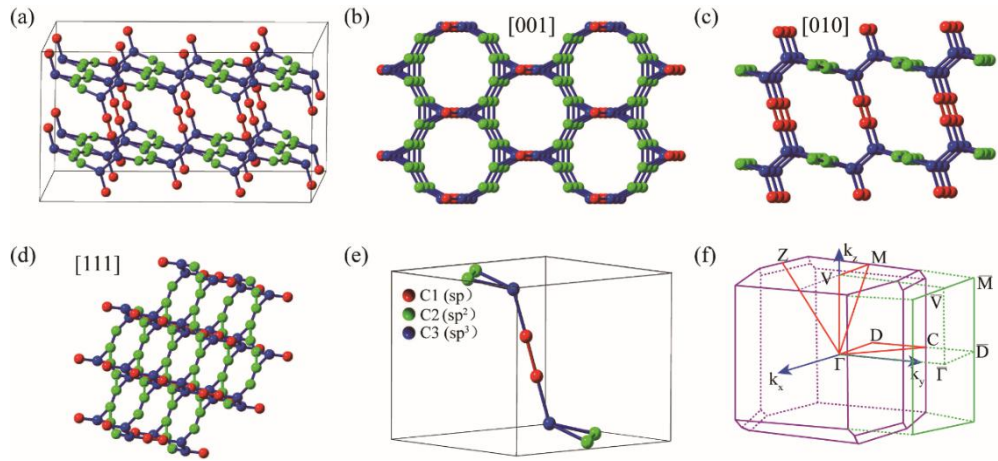

Figure 1. (Color online) (a) The geometrical structure of carboneyane, and views from (b) [001], (c) [010] and (d) [111] directions. (e) The primitive cell of carboneyane, where the carbon atoms are classified into three types C1, C2, C3 by bonding nature, and the red, green and blue colored bonds represent triple, double and single bonds, respectively. (f) The Brillouin zone (BZ) with high symmetry paths (red lines), as well as the projected surface BZ of (010) surface (green rectangle).

The geometric structure of carboneyane is shown in Fig. 1. It adopts a base centered monoclinic lattice with the space group No. 12 (C2/m), with two symmetries:

the inversion symmetry $\mathcal{P}$ and the mirror symmetry $\mathcal{M}_y$. The primitive cell contains 8 C atoms, where three unit vectors are $\vec{a}_1$=(a/2, -b/2, 0), $\vec{a}_2$=(a/2, b/2, 0), $\vec{a}_3$=(c × cosβ, 0, c × sinβ) with a=8.82 Å, b=5.09 Å, c=4.96 Å and b=91.49°. The carbon atoms in carboneyane are classified into three types by bonding nature, where C1, C2, C3 represent *sp*, *sp²* and *sp³* hybridized carbon atoms colored by red, green and blue, respectively, and the ratio of the numbers of *sp*, *sp²* and *sp³* hybridized carbon atoms is 1:2:1.

| Structure | ρ (g/cm³) | d (Å) | $E_{coh}$ (eV/atom) | B (10²GPa) | bonds |
|---|---|---|---|---|---|
| c-diamond | 3.52 | 1.544 | 7.76 | 4.64 | sp³ |
| h-diamond | 3.52 | 1.539, 1.561 | 7.73 | 4.30 | sp³ |
| graphite | 2.22 | 1.422 | 7.89 | 2.94 | sp² |
| T-carbon | 1.50 | 1.502, 1.417 | 6.57 | 1.69 | sp³ |
| graphdiyne | | 1.233-1.432 | 7.11 | | sp-sp² |
| carboneyane | 1.43 | 1.218-1.544 | 6.92 | 1.40 | sp-sp²-sp³ |

Table 1. The equilibrium density (ρ), bond length (d), cohesive energy ($E_{coh}$), bulk modulus (B), and chemical bonds of cubic and hexagonal diamond, graphite, T-carbon, graphdiyne and carboneyane.

The equilibrium density (ρ), bond length (d), cohesive energy ($E_{coh}$), bulk modulus (B), and hybridized chemical bonds of cubic and hexagonal diamond [24], graphite, T-carbon [7], graphdiyne [6] and carboneyane are summarized in Table 1 for comparison. Note that all data presented here were obtained by ourselves using the same method as carboneyane, which can thus be compared reasonably at the same level.

Our calculations indicate that the cohesive energy per atom of carboneyane is 6.92 eV, around 0.35 eV/atom higher than T-carbon at the GGA level, suggesting that this structure could be synthesized. Because of its fluffy structure, carboneyane possesses the smallest equilibrium density and bulk modulus among diamond, graphite and T-carbon.

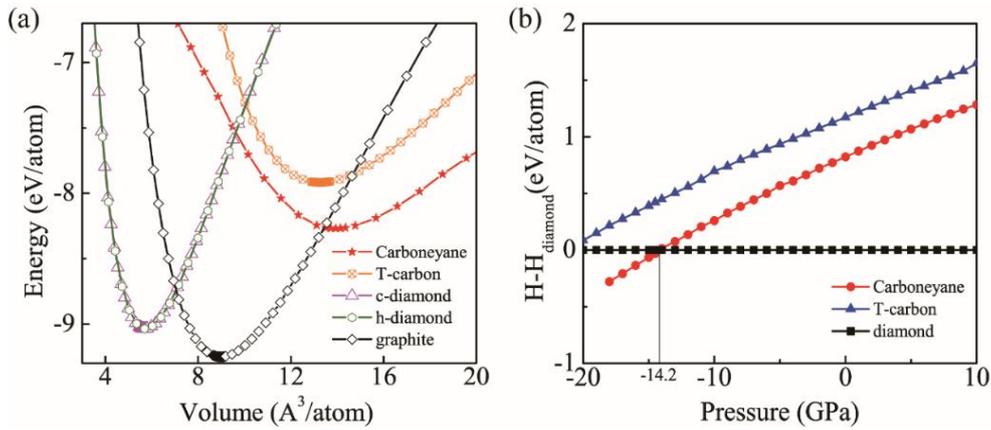

Figure 2. (a) The total energy per atom as function of volume per atom for carboneyane, T-carbon, cubic diamond (c-diamond), hexagonal diamond (h-diamond) and graphite. (b) The enthalpy (H) per atom of carboneyane and T-carbon relative to diamond (Hdiamond) as function of pressure.

The geometric structure of carboneyane is energetically stable, and the minimum energy per atom is -8.27 eV, which is lower than T-carbon but higher than other carbon allotropes, indicating it is a thermodynamically metastable phase, as shown in Fig. 2(a). By calculating the enthalpy of carboneyane, as given in Fig. 2(b), it is obvious that carboneyane possesses an enthalpy lower than T-carbon, indicating it is more stable and can be formed easier than T-carbon. Beyond negative 14.2 GPa, it appears that carboneyane is more stable than diamond in the sense that the relative enthalpy becomes less than zero in this case. To confirm the kinetic stability of carboneyane, its phonon spectra and density of states (DOS) were calculated, as presented in Fig. 3. The obtained phonon eigenvalues can be well explained by considering the bonding nature of the single, double and triple carbon-carbon bonds in this allotrope. The vibrational modes due to the triple yne-bonds can be observed around 2200 cm$^{-1}$ with the carbon-carbon bond length of 1.218 Å (C1-C1), the vibrational modes due to the double ene-bonds are distributed around 1700 cm$^{-1}$ with the carbon-carbon bond length of 1.355 Å (C2-C2), and the vibrational modes due to the single ane-bonds are distributed around 1050 cm$^{-1}$ with the carbon-carbon bond length more than 1.36 Å (C3-C3). The combination modes of these three types of chemical bonds of carbon atoms can be seen below 800 cm$^{-1}$. No negative frequency phonon is observed in the whole BZ, indicating that carboneyane is kinetically stable. To further examine the thermal stability, we performed ab initio molecular dynamics simulations using a 2×2×2 supercell containing 128 carbon atoms. After being heated at 500 K and 1000 K for 9 ps with a time step of 3 fs, no structural changes occur as shown in Fig. S1 (Supplemental Material), revealing that carboneyane is viable for experimental synthesis.

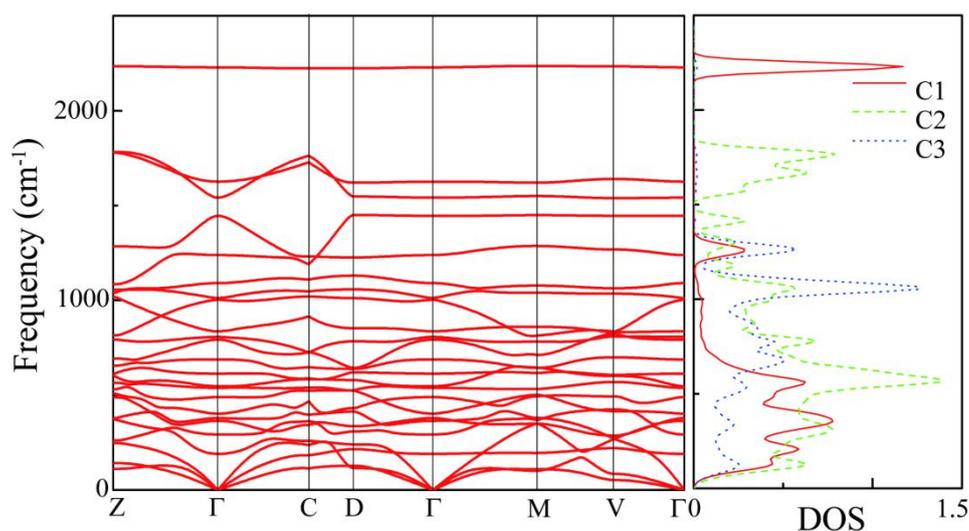

Figure 3. (a) Unit cell and the corresponding BZ of caboneyane. (b) Phonon band structures and phonon partial density of states (DOS) of carboneyane.

To assess the mechanical properties of carboneyane, its elastic constants were calculated, which gives $C_{11}$, $C_{22}$, $C_{33}$, $C_{44}$, $C_{55}$, $C_{66}$, $C_{12}$, $C_{13}$, $C_{23}$, $C_{15}$, $C_{25}$, $C_{35}$ and $C_{46}$ 236.4, 447.8, 426, 95.2, 82, 20.1, 86.8, 58.4, 7.9, 34.9, 26.2, 82.7 and 22.1 GPa, respectively. These values meet the criteria of mechanical stability for monoclinic phase [25]. In terms of the Voigt-Reuss-Hill approximations [26], we can acquire the bulk $B$ and shear $G$ modulus of carboneyane, say 140 and 74 GPa, respectively. The calculated ratio B/G is 1.89, indicating that carboneyane is ductile, because a high (low) B/G value is associated with ductility (brittleness), and the critical value that separates ductile and brittlematerials is about 1.75 according to Pugh's rule [27]. Young's modulus E and Possion's ratio n of carboneyane are obtained by $E = 9GB/(3B+G) = 189$ GPa, $\nu = (3B-2G)/[2(3B+G)] = 0.275$. The smaller Poisson's ratio compared with 0.318 of T-carbon [7] indicates that the bonding is more directional in carboneyane.

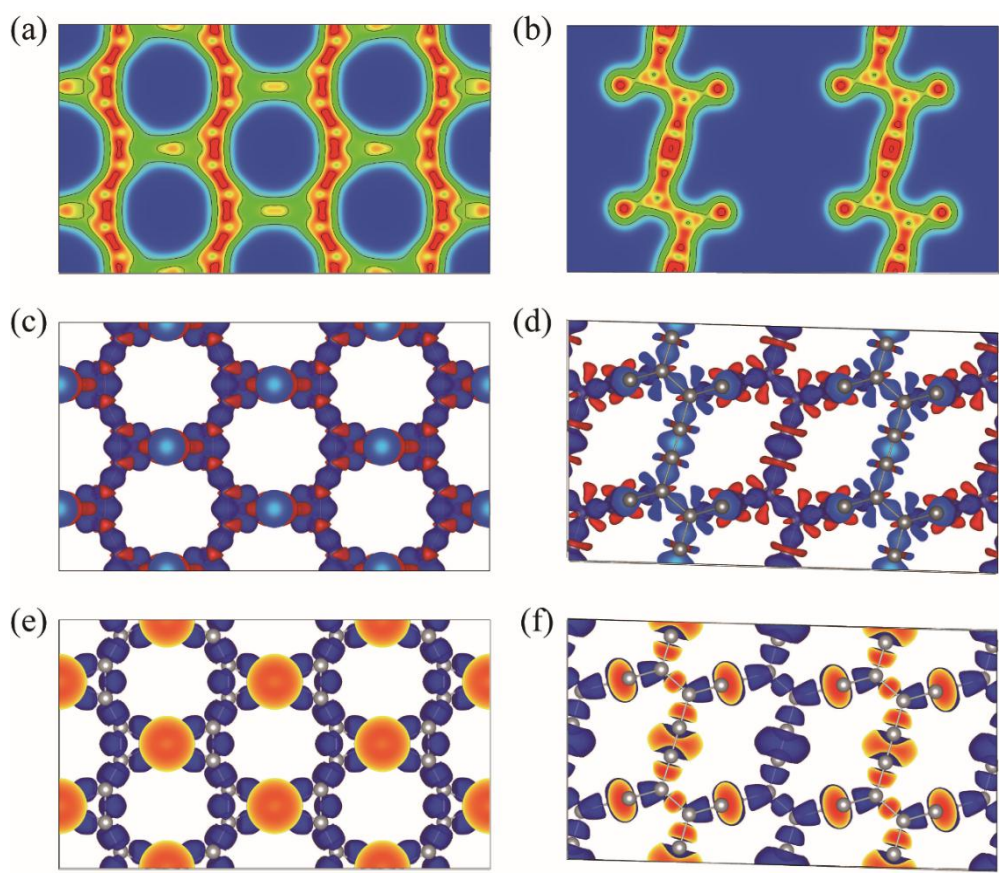

Figure 4. (Color online) Electron charge density, electron density difference and electron localization function maps of carboneyane. The electron charge density maps on (a) (001) and (b) (010) planes, where the warmer (red) colors represent higher local charge density. The electron density difference (EDD) with an isodensity of 0.12 e/Å3 on (c) (001) and (d) (010) planes, respectively. The blue color denotes a gain and the red color a loss of electron density. The electron localization function (ELF) with an isosurface level of 0.7 on (e) (001) and (f) (010) planes.

To better understand the bonding nature of electrons in carboneyane, the electron charge density, electron density difference (EDD) and electron localization function (ELF) maps are illustrated in Fig. 4. The EDD maps reflect the variation of electron density in terms of chemical bonding, as an EDD is plotted by subtracting the overlapping atomic electron density from the self-consistent electron density of carboneyane. The ELF maps give a clear and quantitative description on the basic chemical bonds [high ELF values (0.5 < ELF < 1) indicate the formation of covalent bonds [28,29,30]. It is instructive to note from Figs. 4(a) and 4(b) that electrons in carboneyane are localized on (001) plane along the armchair (y-axis) direction including $sp^2$ carbon atoms and on (010) plane along the zigzag (z-axis) direction containing $sp$ and $sp^3$ carbon atoms. From the EDD maps shown in Figs. 4(c) and 4(d), we can see that there is a larger gain of electron density between triple bonded carbons. Meanwhile, from the ELF maps shown in Figs. 4(e) and 4(f), we find that there is an enhanced localization between triple bonded carbons clearly, which is larger than the localization between the single and double bonded carbons. These results show that the bonding strength of the hybridized bonds $sp$, $sp^2$ and $sp^3$ in carboneyane are different, where the $sp$ bond is the strongest while both $sp^2$ and $sp^3$ bonds are comparable.

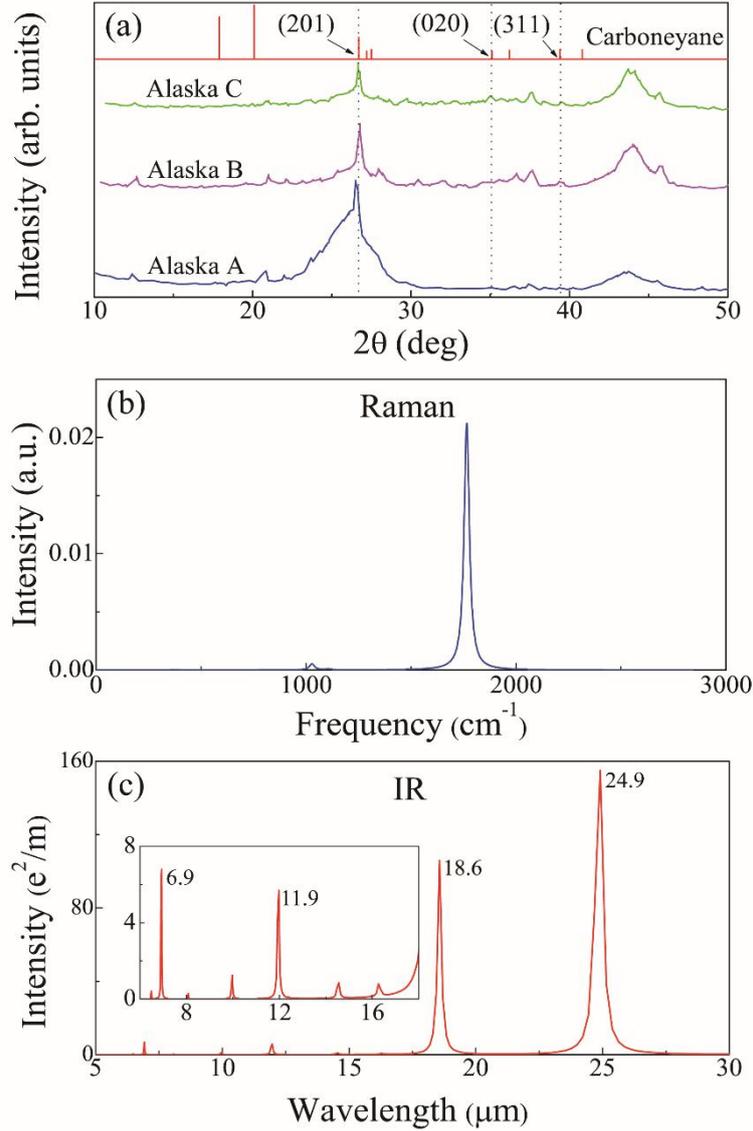

Figure 5. (Color online) (a) The simulated XRD spectra of carboneyane, which is compared with the experimental XRD pattern in Ref. [31]. The used X-ray wavelength is 1.54 Å, as employed in the experiment [31]. (b) The Raman spectra and (c) the infrared spectra of carboneyane.

To provide more information for possible experimental identification, we also simulated the x-ray diffraction (XRD) spectra of carboneyane with wavelength 1.54 Å. The results are presented in Fig. 5(a). The XRD peaks appear at the angles 2θ = 17.9° of (001), 20.1° of (110), 26.7° of (201), 27.2° of (111), 27.5° of (201), 35.1° of (020),

36.2° of (002), 39.4° of (311) and 40.8° of (220) planes. It is interesting that in detonation soot samples Alaska A, B and C [31] with different carbon concentrations, the most remarkable feature of the experimental XRD spectra is the peak around 26.7° which well matches one of calculated XRD peaks of carboneyane. Besides, the experimental peaks around 35.1° and 39.4° of the detonation soot also meet with the calculated data of carboneyane. This observation suggests that carboneyane might be a possible candidate of the carbon phase observed in the detonation soot.

The simulated Raman and infrared (IR) vibrational modes with corresponding frequencies are presented in Figs. 5(b) and 5(c), respectively. The Raman spectra exhibit a well-marked peak at 1760 cm$^{-1}$. The IR spectra show a number of peaks within the wavelength of 5 μm and 30 μm. These attainable features may be useful for experimentally identifying carboneyane in future.

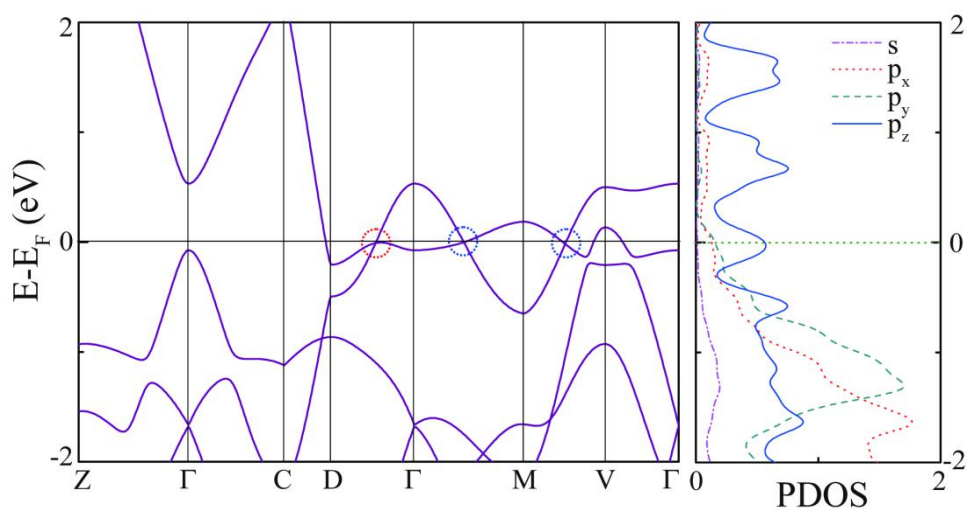

Figure 6. (Color online) Electronic band structures and partial density of states (PDOS) of carboneyane. The blue and red colored circles represent type-I and type-II band dispersions, respectively. The conduction and valence bands at type-I Weyl points have the same sign.

The band structures and partial DOS (PDOS) calculations within the GGA are shown in Fig. 6. In the band structure, there appear linear band crossing points along *D-Γ*, *Γ-M* and *M-V* paths very close to the Fermi level, with both type-I and type-II dispersion marked by blue and red colored circles, respectively. Since the system preserves both $\mathcal{P}$ and $\mathcal{T}$ symmetries, indicating that such linear crossing point cannot be isolated [32,33]. Indeed, a careful scan of the band structure reveals that the crossing between the two bands forms a pair of nodal lines, as illustrated in Figs. 6(a) and 6(b). The two nodal lines are lying in the (010) plane, as protected by the My symmetry. It is interesting that the nodal line here is hybrid type due to the coexistence of both type-I and type-II dispersion. A nodal line can be classified as type-I, type-II [34], or hybrid nodal line [34,35] based on the type of dispersion around the line. A type-I (type-II) nodal line consists solely of nodal points with type-I (type-II) dispersion, whereas a hybrid line contains both. It is important to note that the $k_y = 0$ plane is mirror-invariant, such that the nodal lines are protected by the mirror symmetry $\mathcal{M}_y$, which has been confirmed by our first-principles calculations that the two crossing bands have opposite mirror eigenvalues. The spin-orbit coupling (SOC), which opens up a small gap about 0.1 meV at the band crossing points according to our calculations, is negligibly weak, and thus does not alter the topological nature of carboneyane.

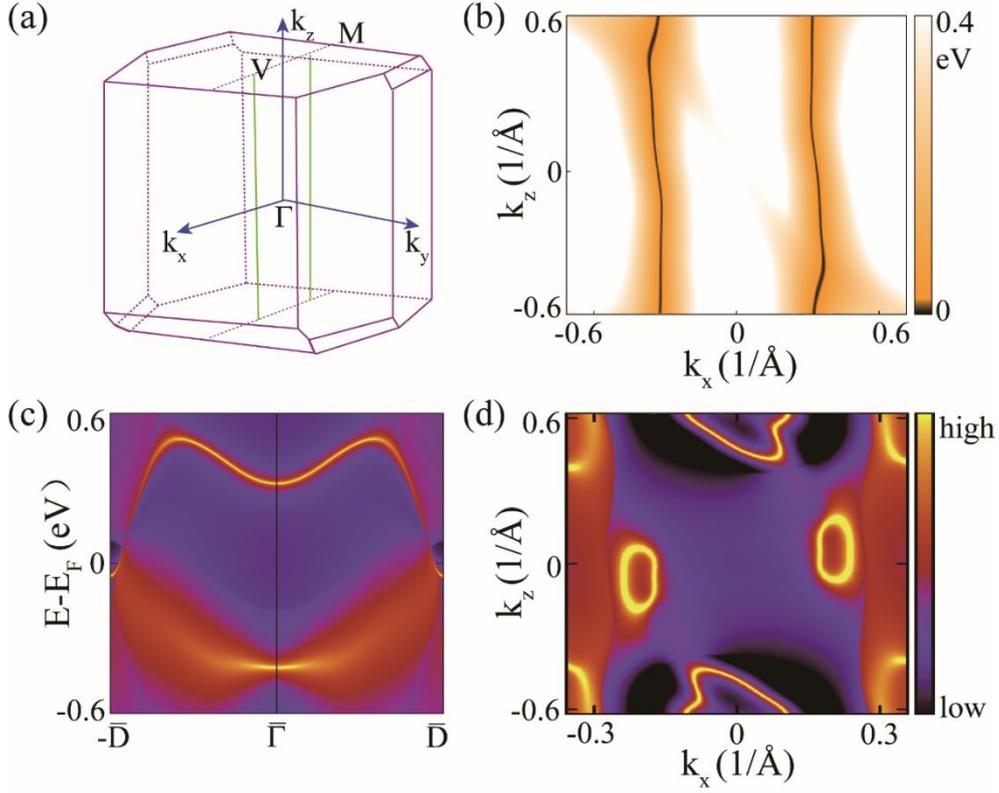

Figure 7. (Color online) (a) Schematic figure showing the two nodal lines (green lines) in the 3D BZ. (b) Shapes of the two Weyl loops obtained from DFT calculations. The color map indicates the local gap between the two crossing bands. (c) Projected spectrum on the (010) surface for carboneyane. (d) A constant energy slice at 0.16 eV, where the bright lines showing the drumhead surface states.

The surface of a nodal line semimetal features drumhead like states. Here, "drumhead" means that these surface states are located in a region bounded by the surface projection of the bulk nodal lines in the surface BZ. In Fig. 7(c), we show the surface states of carboneyane on the (010) surface. Indeed, one observes the drumhead surface bands that emanate from the bulk nodal points, which connects the two nodal lines through the surface BZ boundary. Beside the surface bands from these two points in Fig. 7(c), there exist such bands connecting any two points at the two lines, such that a drumhead surface state emerges inside the projected lines. In Fig. 7(d), we plot the constant energy slice at $E = 0.16$ eV, which cuts through the drumhead, forming two Fermi circles and two arcs, because the drumhead is not completely flat in energy.

From the PDOS, we can observe that the states around the Fermi level come mainly from the $p_z$ orbitals, which determine the metallic nature of carboneyane. Via a careful analysis on PDOS near the Fermi level (-0.3-0.3 eV) [Fig. S2(a)], we uncover that the $p_z$ orbitals near the Fermi level are from p-electron of $sp^2$ hybridized carbon atoms on the (001) plane, giving rise to the conductive property of caboneyane. By computing the conductivity of carboneyane with the package BoltzTrap [36], we find that the conductivity $\sigma_{yy}/\tau$ is about 40 times of $\sigma_{xx}/\tau$ and about 400 times of $\sigma_{zz}/\tau$ with $\tau$ the relaxation time of electrons, indicating a strong anisotropic conductivity [Fig. S2(b)].

Carboneyane possesses a unique fluffy structure with large interspace between carbon atoms, making it easy to hold atoms for adsorption and migration. We investigated the adsorption and diffusion of alkali (Li, Na, K) and alkaline earth (Mg) atoms in carboneyane. The results are collected in Table S1. From Table S1 one may find that the maximal specific capacity was estimated to be 588 $mAhg^{-1}$ for Li, K, Mg atoms, which is larger than that in graphite (372 $mAhg^{-1}$). In addition, the doped Li structure exhibits potential as an ideal anode electrode, which has a remarkable ion migrating barrier of only 0.094 eV much lower than 0.327 eV in graphite. Carboneyane also exhibits a very low voltage, high storage capacity and small volume deformation for Mg ion. Thus, carboneyane could be a good storage and adsorption material for Li, K and Mg atoms, and a promising anode material for lithium-ion and magnesium-ion batteries.

## 4. Conclusion

In summary, by means of first-principle calculations we proposed for the first time a novel carbon allotrope dubbed as carboneyane which contains simultaneously *sp*, *sp²* and *sp³* hybridized chemical bonds in one structure. This carbon allotrope, with a ductile and directional structure, is the lightest among the known three-dimensional carbon allotropes. The calculations on electronic properties of carboneyane show that it is a topological metal with hybrid nodal lines protected by a mirror symmetry near the Fermi level. The XRD spectra show that three peaks of carboneyane well fit with the detonation soot. Upon being obtained, the intriguing carboneyane may be a good storage and adsorption material for Li, K and Mg atoms as well as a promising anode material for lithium-ion and magnesium-ion batteries.

**Conflicts of interest:**

The authors declare no conflicts.

**Acknowledgements**

All calculations are performed on Era at the Supercomputing Center of Chinese Academy of Sciences, MagicCube II in Shanghai Supercomputer center, and Tianhe-2 at National Supercomputing Center in Guangzhou, China. This work is supported in part by the National Key R&D Program of China (Grant No. 2018YFA0305800), the Strategic Priority Research Program of CAS (Grant Nos. XDB28000000,

XBD07010100), and the NSFC (Grant No. 11834014, 14474279, 11504013).

**Appendix**

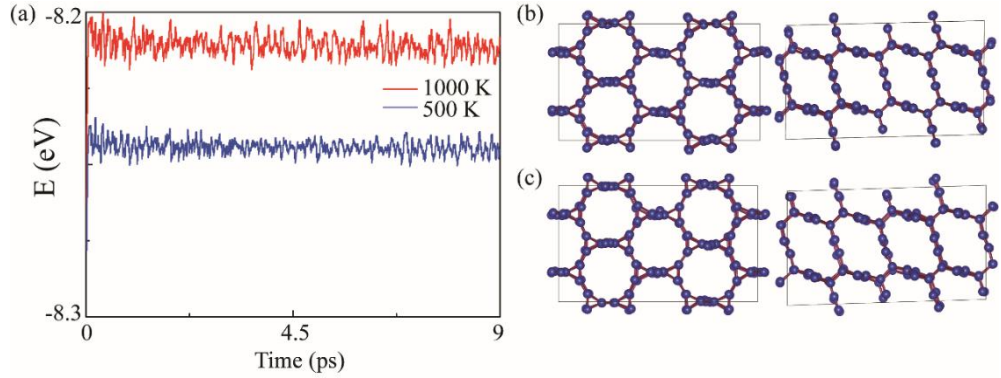

Figure S1. (a) Fluctuation of energy per carbon atom of the supercell containing 128 atoms of carboneyane as a function of time from the molecular-dynamic (MD) simulation at 500 K (blue lines) and 1000 K (red lines). The top and side view of the final crystal structures at (b) 500 K and (c) 1000 K.

From the MD simulation as shown in Fig. S1, one can see that the carboneyane can be stable at 500 K and 1000 K.

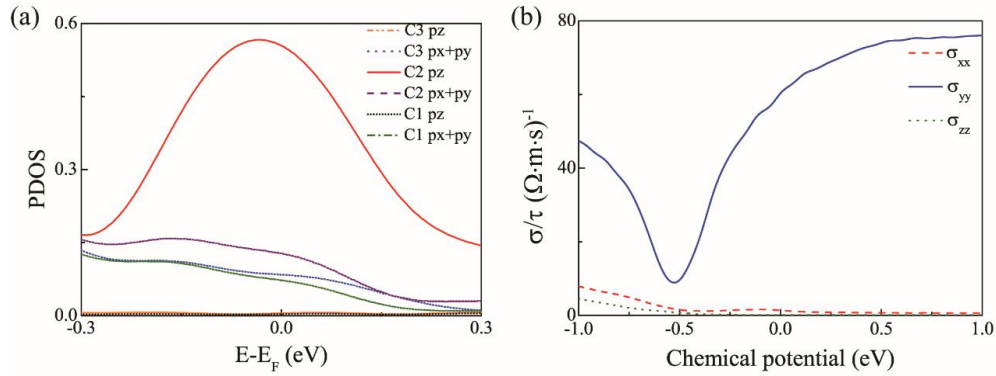

Figure S2. (a) The partial DOS of different hybridized carbon atoms near the Fermi level (-0.3-0.3 eV). (b) The conductivity of carboneyane, where $\sigma_{xx}$, $\sigma_{yy}$ and $\sigma_{zz}$ are colored in red, blue and green, respectively. $\tau$ is the relaxation time of electrons in carboneyane.

Figure S2(a) shows the main contribution to the DOS of electrons near the Fermi level comes from $p_z$ orbital of $sp^2$ hybridized carbon atoms (C2). C2 carbon atoms distributing along the armchair (y axis) direction on (001) plane gives rise to the conductive property of carboneyane. The conductivity $\sigma_{yy}$ is verified to be much larger than $\sigma_{xx}$ and $\sigma_{zz}$ as shown in Fig. S2(b), showing that the conductivity of carboneyane is very anisotropic and directional.

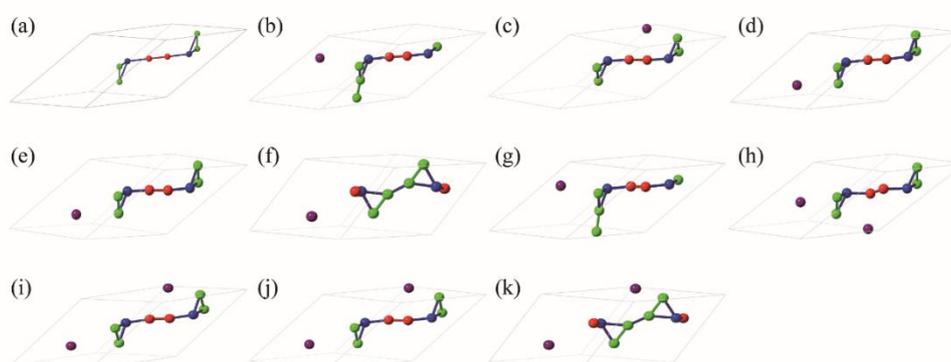

Figure S3. (a) The primitive cell of carboneyane containing eight carbon atoms. The primitive cell of carboneyane doped with one (b) Li, (c) Na, (d) K, (e) Rb, (f) Cs, (g) Mg atom and two (h) Li, (i) K, (j) Rb, (k) Cs atoms. The red, green and blue balls represent C1, C2, and C3 with $sp$, $sp^2$ and $sp^3$ chemical bonds, respectively. The purple balls represent doped metal atoms.

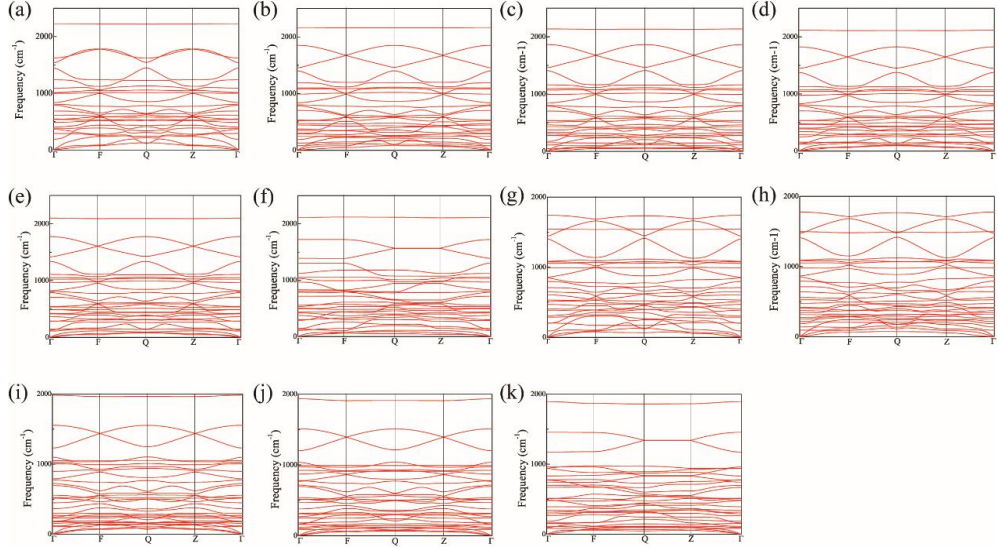

Figure S4. Phonon band structures based on 2 × 2 × 2 supercell of the undoped and doped carboneyane structures with metal atoms corresponding to Fig. S3.

| Structure | Li | Na | K | Mg |
|---|---|---|---|---|
| $C$ (mAh/g) | 558 | 279 | 558 | 558 |
| $D$ (%) | 5.3 | 2.7 | 6.2 | 1.5 |
| $E_b$ (eV) | 2.802 | 2.321 | 2.442 | 1.942 |
| $U$ (V) | 0.94 | 1.07 | 1.37 | 0.40 |
| $E_a$ (eV) | 0.094 | 1.550 | 1.689 | 0.708 |

Table S1. The specific capacity ($C$), volume deformation ($D$), binding energy ($E_b$) between M atom and carboneyane, voltage ($U$) relative to the M metal anode and energy barrier ($E_a$) for M ion migration in carboneyane.

Because of its fluffy structure and all hybridized chemical bonds of carbon atoms, carboneyane can be doped with a number of metal atoms. Here we present a few of them as shown in Fig. S3 for examples. No imaginary frequencies (Fig. S4) were observed through the entire phonon band structures for all the doped structures we have listed, confirming the kinetic stabilities of these structures. We investigated the adsorption and diffusion of alkali (Li, Na, K) and alkaline earth (Mg) atoms in carboneyane. The maximal M (Li, Na, K, Mg) specific capacity ($C$) of carboneyane, volume deformation (*D*) at the maximal specific capacity, binding energy ($E_b$) between M atom and carboneyane, voltage (*U*) relative to the M metal anode, and energy barrier ($E_a$) for M ion migration in carboneyane were calculated, where the specific capacity is defined by $C = nF/m$ with *n* the number of electrons involved in the electrochemical process, *F* the Faraday constant with a value of 26.8 *Ah/mol*, and *m* the molar mass of carboneyane, and the binding energy was calculated by $E_b = E_{C_8} + xE_M - E_{M_xC_8}$ with $M_xC_8$ a primitive cell of carboneyane doped with *x* M atoms. The results are collected in Table S1. From Table S1 one may find that the maximal specific capacity was estimated to be 588 *mAhg*$^{-1}$ for Li, K, Mg atoms, which is larger than that in graphite (372 *mAhg*$^{-1}$). In addition, the doped Li structure (LiC$_4$) exhibits potential as an ideal anode electrode, which has a remarkable ion migrating barrier of only 0.094 eV much lower than 0.327 eV in graphite. Carboneyane also exhibits a very low voltage, high storage capacity and small volume deformation for Mg ion. Thus, carboneyane could be a good storage and adsorption material for Li, K and Mg atoms, and a promising anode material for lithium-ion and magnesium-ion batteries.

| Atom | x | y | z |
|---|---|---|---|
| C1$_1$ | 0.480835 | 0.480835 | 0.380980 |
| C1$_2$ | 0.519165 | 0.519165 | 0.619020 |
| C2$_1$ | 0.153090 | 0.418172 | 0.017015 |
| C2$_2$ | 0.418172 | 0.153090 | 0.017015 |
| C2$_3$ | 0.846910 | 0.581828 | 0.982985 |
| C2$_4$ | 0.581828 | 0.846910 | 0.982985 |
| C3$_1$ | 0.434608 | 0.434608 | 0.100327 |
| C3$_2$ | 0.565392 | 0.565392 | 0.899673 |

Table S2. The atom fractional coordinates of carboneyane.

**References**

[1] Kroto HW, Heath JR, O'Brien, Curl RF, Smalley RE. C60: Buckminsterfullerene. Nature

1985;318:162.

[2] Iijima S. Helical microtubules of graphitic carbon. Nature 1991;354:56,.

[3] Novoselov KS. Electric Field Effect in Atomically Thin Carbon Films. Science 2004;306: 666.

[4] Hoffmann R, Kabanov AA, Golov AA, Proserpio DM. Homo Citans and Carbon Allotropes: For an Ethics of Citation. Angew Chem Int Ed 2016;55:10962 .

[5] Pan B, Xiao J, Li J, Liu P, Wang C, Yang G. Carbyne with finite length: The one-dimensional sp carbon. Sci Adv 2015;1:e1500857.

[6] Li G, Li Y, Liu H, Guo Y, Li Y, Zhu D. Architecture of graphdiyne nanoscale films. Chem Commun 2010;46:3256 .

[7] Sheng X-L, Yan Q-B, Ye F, Zheng Q-R, Su G. T-Carbon: A Novel Carbon Allotrope. Phys Rev Lett 2011;106:155703.

[8] Zhang J, Wang R, Zhu X, Pan A, Han C, Li X, et al. Pseudo-topotactic conversion of carbon nanotubes to T-carbon nanowires under picosecond laser irradiation in methanol. Nat Commun 2017;8:683.

[9] Sheng X-L, Cui H-J, Ye F, Yan Q-B, Zheng Q-R, Su G. Octagraphene as a versatile carbon atomic sheet for novel nanotubes, unconventional fullerenes, and hydrogen storage. J Appl Phys 2012;112:074315.

[10] Li Q, Ma Y, Oganov AR, Wang H, Wang H, Xu Y, et al. Superhard Monoclinic Polymorph of Carbon. Phys Rev Lett 2009;102:175506.

[11] Jo JY, Kim BG. Carbon allotropes with triple bond predicted by first-principle calculation: Triple bond modified diamond andT-carbon. Phys Rev B 2012;86:075151.

[12] Wang J-T, Chen C, Mizuseki H, Kawazoe Y. New carbon allotropes in sp-sp3 bonding networks consisting of C8 cubes. Phys Chem Chem Phys 2018;20:7962.

[13] Feng X, Wu Q, Cheng Y, Wen B, Wang Q, Kawazoe Y, et al. Monoclinic C16: sp-sp hybridized nodal-line semimetal protected by PT-symmetry. Carbon 2018;127:527.

[14] Delodovici F, Manini N, Wittman RS, Choi DS, Fahim MA, Burchfield LA. Protomene: A new carbon allotrope. Carbon 2018;126:574.

[15] Kresse G, Joubert D. From ultrasoft pseudopotentials to the projector augmented-wave method. Phys Rev B 1999;59:1758.

[16] Kresse G, Furthmüller J. Efficient iterative schemes forab initiototal-energy calculations using a plane-wave basis set. Phys Rev B 1996;54:11169.

[17] Kresse G, Hafner J. Ab initiomolecular dynamics for open-shell transition metals. Phys Rev B 1993;48:13115.

[18] Perdew JP, Burke K, Ernzerhof M. Generalized Gradient Approximation Made Simple. Phys Rev Lett 1996;77:3865.

[19] Togo A, Tanaka I. First principles phonon calculations in materials science. Scr Mater 2015 108:1.

[20] Mostofi AA, Yates JR, Pizzi G, Lee Y-S, Souza I, Vanderbilt D, et al. An updated version of wannier90: A tool for obtaining maximally-localised Wannier functions. Comput Phys Commun 2014;185:2309.

[21] Kong X, Li L, Leenaerts O, Liu X-J, Peeters FcoM. New group-V elemental bilayers: A tunable structure model with four-, six-, and eight-atom rings. Phys Rev B 2017;96:035123.

[22] Sancho MPL, Sancho JML, Sancho JML, Rubio J. Highly convergent schemes for the calculation of bulk and surface Green functions. J Phys F: Met Phys 1985;15:851.

[23] Wu Q, Zhang S, Song H-F, Troyer M, Soluyanov AA. WannierTools: An open-source software package for novel topological materials. Comput Phys Commun 2018;224:405.
[24] Mao WL, "Bonding Changes in Compressed Superhard Graphite. Science 2003;302:425.
[25] Wu Z-j, Zhao E-j, Xiang H-p, Hao X-f, Liu X-j, Meng J. Crystal structures and elastic properties of superhard IrN2 and IrN3 from first principles. Phys Rev B 2007;76:054115.
[26] Hill R. The Elastic Behaviour of a Crystalline Aggregate. Proc Phys Soc London 1952 Sect A;65:349,.
[27] Pugh SF. XCII. Relations between the elastic moduli and the plastic properties of polycrystalline pure metals. Philos Mag 1954;45:823.
[28] Silvi B, Savin A. Classification of chemical bonds based on topological analysis of electron localization functions. Nature 1994;371:683.
[29] Becke AD, Edgecombe KE. A simple measure of electron localization in atomic and molecular systems. J Chem Phys 1990;92;5397.
[30] Savin A, Jepsen O, Flad J, Andersen OK, Preuss H, von Schnering HG. Electron Localization in Solid-State Structures of the Elements: the Diamond Structure. Angew Chem Int Ed in English 1992;31:187.
[31] Pantea D, Brochu S, Thiboutot S, Ampleman G, Scholz G. A morphological investigation of soot produced by the detonation of munitions. Chemosphere 2006;65:821.
[32] Xiao D, Chang M-C, Niu Q. Berry phase effects on electronic properties. Rev Mod Phys 2010;82:1959.
[33] Weng H, Liang Y, Xu Q, Yu R, Fang Z, Dai X, et al. Topological node-line semimetal in three-dimensional graphene networks. Phys Rev B 2015;92:045108.
[34] Li S, Yu Z-M, Liu Y, Guan S, Wang S-S, Zhang X, et al. Type-II nodal loops: Theory and material realization. Phys Rev B 2017;96:081106.
[35] Zhang X, Yu Z-M, Lu Y, Sheng X-L, Yang HY, Yang SA. Hybrid nodal loop metal: Unconventional magnetoresponse and material realization. Phys Rev B 2018;97:125143.
[36] Madsen GKH, Singh DJ. BoltzTraP: A code for calculating band-structure dependent quantities. Comput Phys Commun 2006;175:67.